# Observation of fractal topological states in acoustic metamaterials

Shengjie Zheng[1, *], Xianfeng Man[2, *], Ze-Lin Kong[3, *], Zhi-Kang Lin[3, *], Guiju Duan[1], Ning Chen[1], Dejie Yu[1], Jian-Hua Jiang[3, †], Baizhan Xia[1, †]

[1]State Key Laboratory of Advanced Design and Manufacturing for Vehicle Body, Hunan University, Changsha, Hunan 410082, China

[2]College of Mechanical and Electrical Engineering, Changsha University, Changsha, Hunan 410022, China

[3]Institute of Theoretical and Applied Physics, School of Physical Science and Technology & Collaborative Innovation Center of Suzhou Nano Science and Technology, Soochow University, Suzhou, Jiangsu 215006, China

*These authors contributed equally to this work.

†Corresponding author. Email: jianhuajiang@suda.edu.cn, xiabz2013@hnu.edu.cn.

**Abstract: Topological phases of matter have been extensively investigated in solid state materials and classical wave systems with integer dimensions. However, topological states in non-integer dimensions remain largely unexplored. Fractals, being nearly the same at different scales, are one of the intriguing complex geometries with non-integer dimensions. Here, we demonstrate acoustic Sierpiński fractal topological insulators with unconventional higher-order topological phenomena via consistent theory and experiments. We discover abundant topological edge and corner states emerging in our acoustic systems due to the rich edge and corner boundaries inside the fractals. Interestingly, the numbers of the edge and corner states scale the same as the bulk states with the system size and the exponents coincide with the Hausdorff fractal dimension of the Sierpiński carpet. Furthermore, the emergent corner states exhibit unconventional spectrum and wave patterns. Our study opens a pathway toward topological states in fractal geometries.**

Topological phases of matter that manifest elegantly the geometric properties of the Bloch bands have been a focus of condensed matter physics [1]. Research on topological physics was further extended to photonic [2-9], acoustic [10-18], and other [19-25] metamaterials that are advantageous in revealing novel topological phases (such as Floquet [4, 26-30], higher-order [15-17, 22-25, 31-38] and fragile [39-45] topological phases) and topological phenomena (such as bulk-defect correspondences [46-55]). However, so far, studies are focused only on systems with integer dimensions.

Fractals are prototype candidates toward systems with non-integer dimensions [56]. The study of the interplay between fractal geometry and topological bands is still at its infant stage. Although many fascinating topological phenomena in fractal geometries have been predicted theoretically [57-66], their experimental realizations are, however, still missing. Since fractal topological states are hard to realize in solid state materials within the current technology, photonic and acoustic metamaterials could play an important role in such studies.

Here, we realize fractal higher-order topological states based on the two-dimensional (2D) Su-Schrieffer-Heeger (SSH) model using acoustic metamaterials. By studying the topological boundary states in various generations of the square Sierpiński carpets, we find several unconventional properties of the higher-order topological states in fractal geometries. First, due to their self-similarity properties, Sierpiński carpets have much more edge and corner boundaries than conventional 2D systems. As a consequence, the numbers of the edge and corner states are much larger than those in conventional 2D systems. These numbers even become comparable with the number of the bulk states, leading to the conclusion that in fractals the edge, corner and bulk have the same fractional dimension. For the Sierpiński carpets studied here, they are of exactly the same Hausdorff fractal dimension $d_f = \log(8)/\log(3) \approx 1.893$. Second, as fractals have abundant edge (corner) boundaries, e.g., outer and inner edges (corners), yielding topological edge (corner) states with rich wave patterns. For instance, the inner corner states have wavefunctions distinct from the conventional corner states. We unveil these findings with consistent theory, simulation and experiments which paves the way toward the study of topological states in fractal geometries.

We start with a 2D SSH model building on square Sierpiński carpets. For the 2D SSH model, there are four sites in each unit-cell. We consider here the tight-binding model with only the nearest-neighbor couplings, which already captures the main physics. Here, the 2D SSH model has fourfold rotation symmetry, and the inter-unit-cell couplings ($t_2$) are stronger than the intra-unit-cell couplings ($t_1$) [see Fig. 1(a)]. Putting this model into the square Sierpiński carpets greatly enriches the boundaries, since the Sierpiński carpets are constructed by cutting a square into nine congruent subsquares and removing the central subsquare and repeating this procedure iteratively to the subsquares. For instance, the first-generation Sierpiński carpet G(1) has a hole in the center which serves as additional edge and corner boundaries [Fig. 1(a)]. The second-generation Sierpiński carpet G(2) contains more edge and corner boundaries in its central hole and the holes in the subsquares. Higher generations of Sierpiński carpet lattices are discussed in Supplemental Material [67]. In our construction, each smallest subsquare contains a unit-cell of the 2D SSH model.

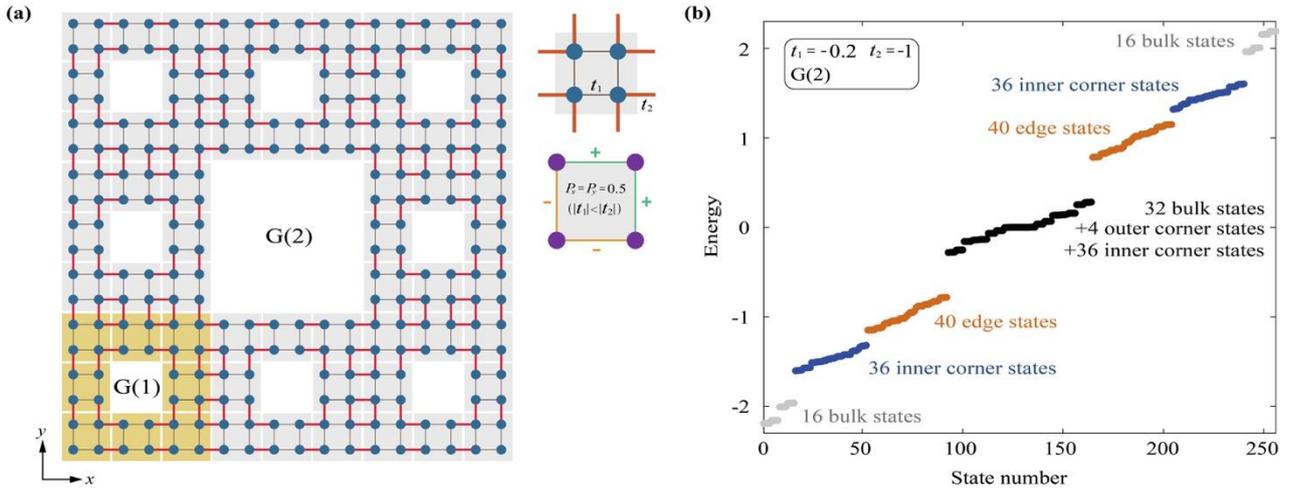

FIG. 1. Sierpiński fractal higher-order topological states. (a) The G(2) Sierpiński fractal tight-binding system based on the 2D SSH model. G(1) is part of the G(2) system. (b) Eigenstates spectrum of the G(2) Sierpiński fractal tight-binding system.

In the conventional 2D geometry, the SSH model is a higher-order topological insulator (HOTI) that hosts topological edge and corner states. The bulk topological invariants are the polarizations along the $x$ and $y$ directions which are determined by the Zak phases,

$$P_i = -\frac{1}{(2\pi)^2} \int_{BZ} A_i(\boldsymbol{k}) d^2\boldsymbol{k}, \ i = x, y. \tag{1}$$

In the region with $|t_1| < |t_2|$, one has $P_x = P_y = 0.5$, whereas for $|t_1| > |t_2|$, one has $P_x = P_y = 0$. We calculate the eigenstates of the fractal SSH model with $t_1 = -0.2$ and $t_2 = -1$. The energy spectrum for the Sierpiński carpet G(2) is given in Fig. 1(b). We find that the numbers of the edge and corner states are extensive: there are 64 bulk states, 80 edge states and 112 corner states. This property, which is a main feature of fractal topological states [57-66], is due to the abundance of the edge and corner boundaries in fractal geometries.

To further analyze this property, we tune the weak coupling $t_1$ to zero. While this process is adiabatic (i.e., the band gap remains open and the topological invariants remain the same), it provides a way to count the bulk, edge, and corner states directly. In this limit, the Sierpiński carpets consist of four types of structure elements, i.e., monomers, dimmers, trimers, and tetramers, that support various local orbitals (Fig. 2). Among them, the monomers emerge at the outer corner boundaries and support the zero-energy outer corner states [see Fig. 2(a)]. Dimmers emerge at the outer and inner edge boundaries and support the edge states localized on these boundaries [see Fig. 2(b)]. A dimer supports a bonding state and an anti-bonding state which form the two edge spectral continua above and below zero energy, respectively. Trimers come from the inner corner boundaries, while tetramers come from the bulk. A trimer supports three local orbitals, one at negative energy (type I), another at zero energy (type II), and the third at positive energy (type III). Type I and III inner corner states are the symmetric eigenstates, while type II inner corner state is the anti-symmetric state [see Fig. 2(c)]. The emergence of the inner corner states is another salient feature of fractal HOTIs. The tetramers give four eigenstates which form three bulk spectral continua at positive, negative and zero energies [see Fig. 2(d)]. We find that the number of monomers ($N_1$), dimers ($N_2$), trimers ($N_3$) and tetramers ($N_4$) are determined by the fractal geometry as follows (see Supplemental Material [67]),

$$N_1 = 4, \quad N_2 = \frac{16}{5} \times 3^n + \frac{1}{35} \times 8^{n+1} - \frac{24}{7}, \tag{2a}$$

$$N_3 = \frac{4 \times (8^n - 1)}{7}, \quad N_4 = \frac{8}{7} - \frac{8}{5} \times 3^n + \frac{16}{35} \times 8^n, \tag{2b}$$

for the G($n$) ($n$=1, 2, 3...) Sierpiński carpet. The above equations dictate the numbers of the outer corner states ($N_{oc} = N_1$), edge states ($N_{ed} = 2N_2$), inner corner states ($N_{ic} = 3N_3$), and bulk states ($N_{bk} = 4N_4$). These numbers satisfy the sum rule, $N_{oc} + N_{ed} + N_{ic} + N_{bk} = 4 \times 8^n$, i.e., the total number of eigenstates is equal to the total number of lattice sites. From the above equations, one finds that the numbers of the edge and corner states scale the same as the bulk states with the system size and the dominant exponents coincide with the Hausdorff fractal dimension of the Sierpiński carpet $d_f = \log(8)/\log(3)$. Particularly, in the limit of $n \to \infty$, the percentages of outer corner states, edge states, inner corner states, and bulk states are 0, 11.43%, 42.86%, and 45.71%, respectively (see Supplemental Material [67]).

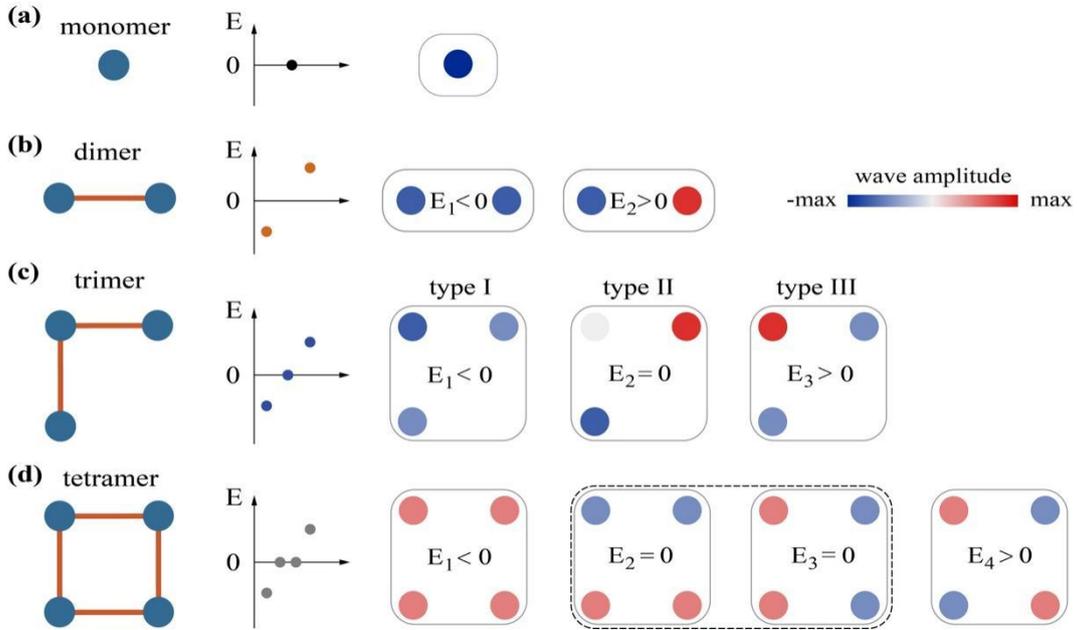

FIG. 2. Basic structure elements and eigenstates for the Sierpiński fractal tight-binding model with $t_1 = 0$. (a) Monomers at the outer corner boundaries. Each monomer supports a zero-energy outer corner state. (b) Dimers at the inner and outer edge boundaries. Each dimer supports a bonding and an anti-bonding edge states. (c) Trimers at the inner corner boundaries. Each trimer supports three inner corner states with negative (type I), zero (type II) and positive (type III) energies, respectively. (d) Tetramers in the bulk. Each tetramer supports a negative energy, a positive energy and two zero energy bulk states.

We now focus on the experimental realization of the fractal HOTI using acoustic metamaterials. For this purpose, we construct an acoustic 2D SSH metamaterial of which the unit-cell structure is given in Fig. 3(a). There are four cylindrical acoustic cavities in each unit-cell. Their mutual couplings are realized by the tubes connecting them. To realize the HOTI phase, we use thick (thin) tubes for the inter-unit-cell (intra-unit-cell) couplings. The geometric parameters are given in the caption of Fig. 3. We calculate the acoustic spectrum of the fractal Sierpiński carpet G(2) using full-wave finite-element simulations. The results presented in Fig. 3(b) show similar spectrum as in Fig. 1(b). The new feature here is that, due to the absence of chiral symmetry in acoustic systems, the outer corner states, the type II inner corner states and the bulk states are spectrally separated. In comparison, in the tight-binding fractal system, these states spectrally overlap with each other. This feature provides advantages in the experimental discernment and detection of these states. In contrast, when the intra-unit-cell couplings are stronger than the inter-unit-cell couplings, the system is trivial and there is no edge or corner state (see Supplemental Material [67]).

The designed G(2) fractal acoustic metamaterial is fabricated using 3D printing technology based on photosensitive resins. We detect the eigenstates via acoustic pump-probe measurements [see Fig. 3(c)] (see details in Supplemental Material [67]). Here, we categorize all acoustic cavities into four different types, those in the bulk region (i.e., the tetramers), those in the edge region (i.e., the dimers), those in the inner corner region (i.e., the trimers), and those in the outer corner region (i.e., the monomers). In each region, we perform the pump-probe measurements for each cavity where we place a small speaker at the bottom of the cavity and a tiny microphone at the top of the same cavity. In such a pump-probe measurement, the measured response spectrum gives approximately the local density of states for the acoustic phonons in each cavity. In Fig. 3(d), we give representative response spectra for the bulk, edge, outer corner, and inner corner regions. We find that the bulk pump-probe response (the gray line) has three peaks at 4742 Hz, 5564 Hz, and 6514 Hz which coincident with the bulk continua in Fig. 3(b). For the edge pump-probe spectrum (the green line), there are two peaks at the frequencies of 5003 Hz and 5893 Hz, which are consistent with the two edge continua in Fig. 3(b). For the outer corner pump-probe spectrum (the red line), there is only one peak at the frequency of 5287 Hz, which is consistent with the calculated eigen-frequency of the outer corner states 5247 Hz.

Two representative inner corner pump-probe response curves are presented in Fig. 3(d) [the purple and blue curves; the detection cavities are at the purple and blue dots in Fig. 3(c), respectively]. The pump-probe response at the purple dot has only two peaks (at 4896 Hz and 6207 Hz), whereas the pump-probe at the blue dot shows three peaks (at 4896 Hz, 5437 Hz, and 6207 Hz). These differences, in fact, reveal the unique nature of the inner corner states. As shown in Fig. 2(c), the inner corner states consist of three different types: type II (anti-symmetric), type I and III (symmetric). Type II inner corner states have vanishing amplitude at the exact corner (i.e., the purple dot) and thus cannot be probed by the pump-probe response at the purple dot. Therefore, such a pump-probe setup can only probe the types I and III inner corner states, whereas the pump-probe response at the blue dot can probe all three types of the inner corner states.

In addition, the detected acoustic signal (specifically, acoustic pressure amplitude) distributions at the resonances of the outer corner modes (at 5287 Hz), the inner corner modes (at 5437 Hz and 6207 Hz), the edge modes (at 5893 Hz), and the bulk state modes (at 6514 Hz) in Fig. 3(e) show excellent agreement with the features of the eigenstates illustrated in Fig. 2 and those obtained from acoustic eigenstates calculations (see Supplemental Material [67]).

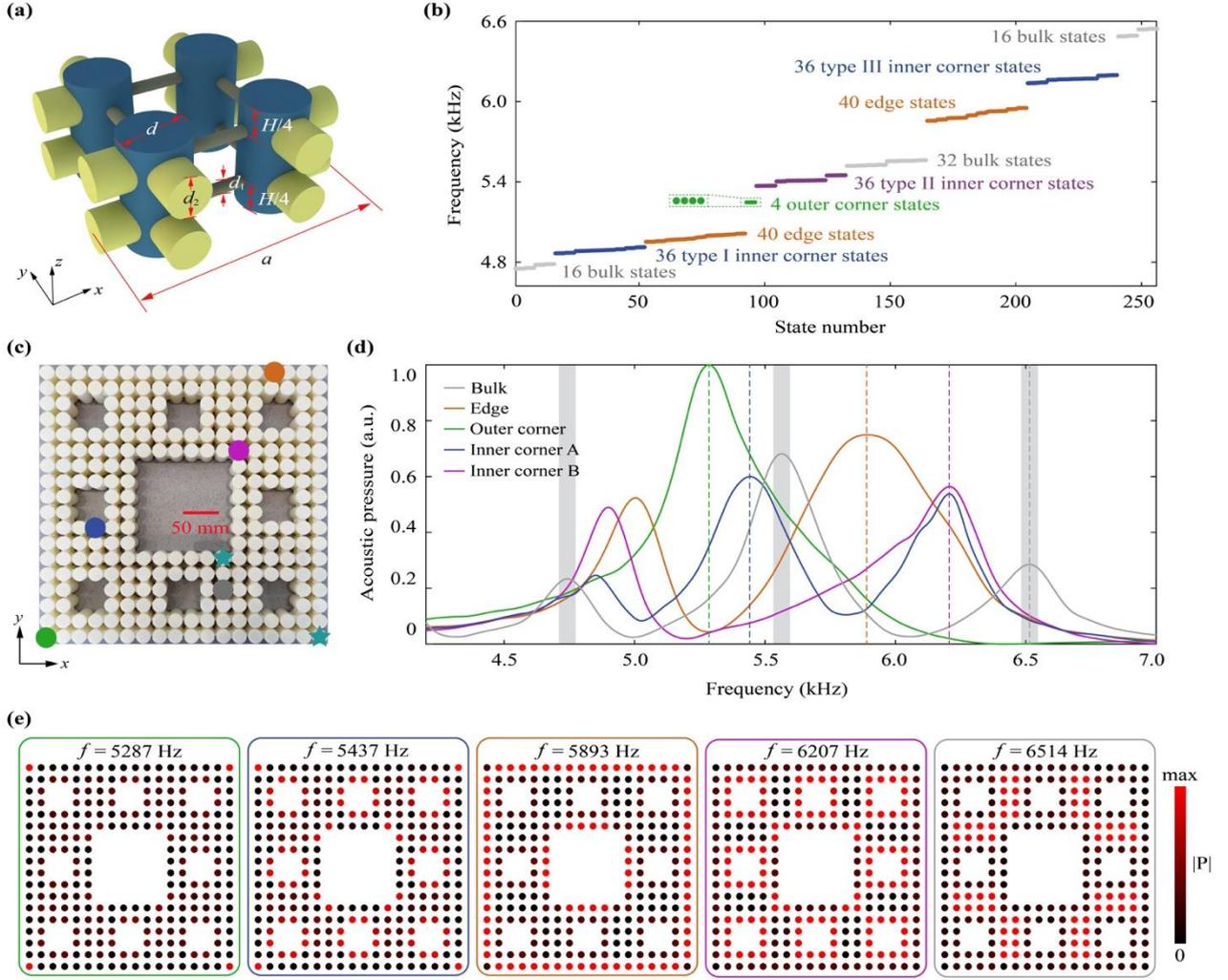

FIG. 3. Experimental observation of Sierpiński fractal higher-order topological states. (a) Unit-cell structure of the acoustic metamaterial. Lattice constant is $a = 58$ mm. $H$=33 mm is the height of the cylindrical cavities. Other parameters are $d$=19 mm, $d_1$=3 mm and $d_2$=12 mm. (b) Calculated eigen-frequencies of the G(2) Sierpiński fractal higher-order topological acoustic metamaterial from full-wave simulations. Different types of eigenstates (labeled by distinct colors) and their numbers are given. (c) Photograph of the fabricated fractal acoustic metamaterial. Green, blue, purple, gray, and brown dots denote the local pump-probe for the outer corner (green), inner corner (blue and purple), bulk (gray), and edge (brown) states. (d) Detected acoustic pressure versus the probing frequency for various local pump-probe configurations to discern different eigenstates. (e) Scanned local pump-probe acoustic pressure (absolute value) at different frequencies corresponding to the resonance peaks of the outer corner states (green box), type II inner corner states (blue box), edge states (brown box), type III inner corner states (purple box), and bulk states (gray box). G(1) and G(3) acoustic metamaterials are discussed in Supplemental Material [67].

To reveal the unique properties of the corner states in the fractal lattice, we experimentally probe the acoustic phase distribution at the outer and inner corner boundaries. For this purpose, we place the acoustic source far away from the detection region to reduce the non-eigenstates (evanescent waves) contributions. For instance, to study the outer corner states, a tiny speaker is inserted into an acoustic cavity from the bottom at another corner boundary [see the right inset of Fig. 4(a), the star labels the source while the dot labels the detector]. The phase profiles of the corner states are probed using a tiny microphone connected to a network analyzer when the excitation frequency sweep across the resonant peak of the concerned corner states.

As shown in Fig. 4(a), the outer corner states are strongly localized at the corner cavity. Meanwhile, due to the dipole nature of all the eigenstates (i.e., all the eigenstates examined here are associated with the lowest $p_z$ orbitals in the cylindrical acoustic cavities), the acoustic phases at the upper and lower halves of the cavity differ by $\pi$. This feature is confirmed by both the simulated outer corner state (the left inset) and the measured acoustic phases in Fig. 4(a), showing consistency with previously observed corner states [31-38].

For the inner corner states, as shown in Figs. 4(b)-4(d), the wave patterns are much different. For instance, in Fig. 4(b), the acoustic phases in the three cavities around the inner corner boundary are nearly the same, which is a signature of the type-I corner states [see the inset and Fig. 2(c)]. Meanwhile, all these cavities have $\pi$ phase difference between the upper and lower part of the cavity, confirming again the $p_z$ orbital nature of the eigenstates. In Fig. 4(c), the acoustic phases at the detection positions 1 and 2 have $\pi$ difference, revealing the anti-symmetric nature of the type II inner corner states (see inset for the eigenstate). In Fig. 4(d), the acoustic phases at the detection positions 1 and 3 are nearly the same, whereas the acoustic phases at the detection positions 1 and 2 have $\pi$ difference, which is consistent with the eigenstate wavefunction (see the inset). Note that results in Figs. 4(b)-4(d) are detected at the same inner corner while only the frequency ranges are tuned around the resonance peak for each type of corner states. These results double confirm the emergence of versatile inner corner states in fractal HOTIs.

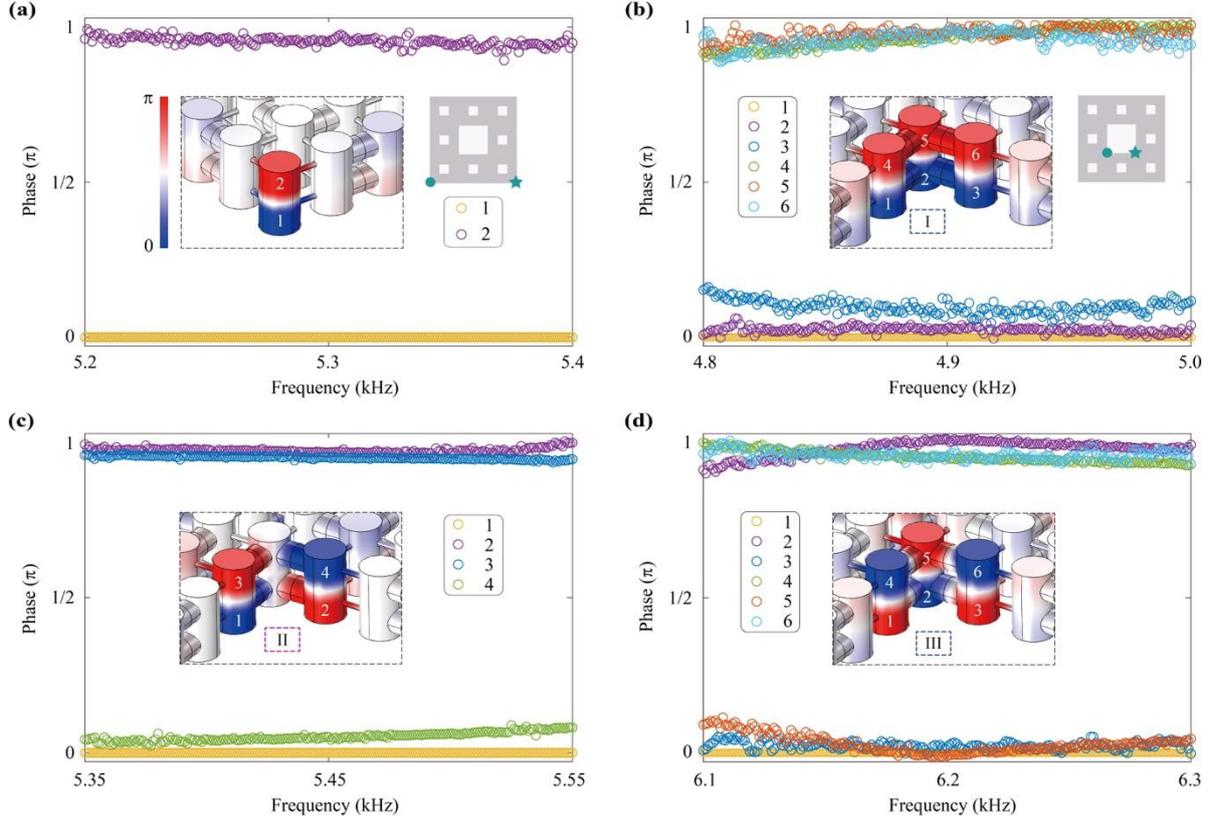

FIG. 4. Detecting the phase profiles of the corner states in fractal lattice G(2). (a) Measured acoustic phases at the detection positions 1 and 2 at an outer corner boundary from 5.2 kHz to 5.4 kHz. (b) Measured acoustic phases at the detection positions 1-6 at an inner corner boundary from 4.8 kHz to 5.0 kHz. In (a) and (b), the left inset: simulated wavefunction of the concerned corner state from eigenstates simulations. Right inset: schematic of the pump-probe setup for the detection of the outer (a) and inner (b) corner states where the star depicts the source and the dot depicts the detector. (c)-(d) Measured acoustic phases at the detection positions 1-6 at the same inner corner boundary in different frequency ranges. Insets: calculated wavefunction of a type II (c) and III (d) inner corner states from eigenstates simulations.

Through extensive acoustic pump-probe measurements with various configurations, we reveal unconventional topological phenomena in fractal lattices. Other Sierpiński carpets with different fractal dimensions are studied in Supplemental Material [67] which give similar phenomena. Our work unveils an intriguing regime where the emergent phenomena directly manifest the fractal geometry in physics. These emergent phenomena are of fundamental importance and potential application values. In particular, the abundant edge and corner states in fractal lattices could provide novel applications when generalized to photonic systems to realize fractal topological lasing that exploits the enhanced local density of states of photonic edge and corner modes.

**Acknowledgements**

This work is supported by the National Natural Science Foundation of China (Grant Nos. 12125504, 12072108, 51621004, 51905162), the Priority Academic Program Development (PAPD) of Jiangsu Higher Education Institutions, and Hunan Provincial Natural Science Foundation of China (Grant No. 2021JJ40626).